\documentstyle[preprint,amsfonts,aps]{revtex}

\begin{document}
\preprint{\vbox{\hbox{Phys. Rev. Lett., in press}
                \hbox{\tt cond-mat/9505100}
                \hbox{}\hbox{}}}

\title{Kondo Effect in a Luttinger Liquid:\\
Exact Results from Conformal Field Theory}

\author{Per Fr\"ojdh$^1$ and Henrik Johannesson$^2$ \\ \phantom{x}}

\address{{}$^1$Department of Physics, University of
Washington, P.O. Box 351560,\\  Seattle, Washington 98195-1560}

\address{{}$^2$Institute of Theoretical Physics, Chalmers University of
Technology and G\"oteborg University, S-412 96 G\"oteborg, Sweden\\}

\maketitle
\vspace{6mm}
\begin{abstract}
We report on exact results for the low-temperature thermodynamics of
a spin-$\frac{1}{2}$ magnetic impurity coupled to a one-dimensional
interacting electron system.
By using boundary conformal field theory, we show
that there are only two types of critical behaviors consistent with
the symmetries of the problem: {\em either} a local Fermi liquid, {\em
or} a theory with an anomalous response identical to that recently proposed
by Furusaki and Nagaosa.
Suppression of back scattering off the impurity leads to the same
critical properties as for the two-channel Kondo effect. 
\end{abstract}

\newpage

Electron correlations often play an important role in
condensed matter of reduced dimensionality. A key issue, raised by
experiments on mesoscopic quantum dots and wires \cite{Timp}, is
how to describe the interplay between impurity and correlation
effects. For electrons in one dimension (1D), it has long been known that any
finite concentration of impurities leads to Anderson localization
\cite{Chiral}, but as shown recently, even a {\em single} potential
scatterer may dramatically influence the physics in the presence of
repulsive e-e interactions: at $T=0$ the scatterer acts as a
perfectly reflecting barrier \cite{KF}.

The case of a {\em dynamical} scatterer, like a magnetic impurity, is
less well understood. The 3D analogue, with noninteracting quasiparticles
representing the electrons {\em (Fermi liquid)}, is that of the Kondo
problem \cite{Hewson}. By symmetry, it can be modeled by a 1D gas of free
chiral particles coupled to a magnetic impurity, allowing for
an exact solution \cite{AW}.
In contrast, the case of fully interacting 1D electrons in the presence of
a magnetic impurity largely remains to be explored.
Laboratory studies of artificial potential defects (``antidots'') in
quantum  wires
have recently been reported \cite{Kirzcenow}, and the experimental study
of an interacting electron system coupled to a {\em spinfull}
defect may soon be within reach. (A possible realization is a quantum dot
containing two spin levels and coupled to two narrow leads \cite{Meir}.) This
poses a challenge to the theorist. In 1D the e-e interaction
removes the single-particle spectrum, and the electrons effectively get
replaced by new collective excitations, separately carrying spin and
charge {\em (Luttinger liquid)} \cite{Haldane}. A magnetic impurity, on
the other hand, couples to individual electrons, and it is {\em a priori}
not clear how to incorporate its description in that of the spin-charge
separated modes.

The problem was recently considered by Furusaki and Nagaosa
\cite{FN}, expanding on earlier work by Lee and Toner \cite{Lee}. These authors
studied a Tomonaga-Luttinger model \cite{Solyom}, with the electrons coupled to
a local magnetic moment via a Kondo exchange. Using ``poor man's scaling,''
an infinite-coupling fixed point was identified, suggesting a completely
screened impurity at low temperatures.
The impurity specific heat, as well as the
conductance, was argued to exhibit an anomalous temperature dependence,
with a leading term $T^{(1/K_{\rho})-1}$, $K_{\rho}$ being the
Luttinger liquid ``charge parameter''
\cite{Haldane}. However, the  validity of the result remains
unclear, as it relies
upon a perturbative expansion in a strong coupling region where perturbation
theory in fact loses its meaning.

In this Letter we study the problem using boundary conformal field theory
$(BCFT)$ \cite{Cardy}. The heart of the method, pioneered by Affleck and Ludwig
\cite{A,AL}, is to replace the impurity by a scale invariant boundary
condition. Combined with the machinery of $BCFT$ this approach has
proven very powerful, and has opened up an entirely new vista on quantum
impurity problems. As with any application of conformal theory,
the method gives a classification of {\em possible} critical behaviors. Being
exact, this information is extremely valuable as it places
strong constraints on any constructive theory of an impurity problem.
In the present case several new features appear, making the
identification of boundary condition less obvious. Still,
by exploiting symmetry arguments we arrive at an exact result showing
that the problem must renormalize to one of only two possible
fixed-point theories.

We describe the electrons by a Tomonaga-Luttinger
Hamiltonian with repulsive interaction $(g > 0)$ \cite{Solyom}:
%%%%%%%%%%%%%%%%%%%%%%%%%%%%%%%%%%%%%%%%%%%
\begin{eqnarray}
{\cal H}_{TL} & = & \frac{1}{2 \pi} \int \! dx \biggl\{  v_F
\biggl[ : \! \psi^{\dagger}_{L,\sigma}(x) i \frac{d}{dx}
\psi^{\mathop{\phantom{\dagger}}}_{L,\sigma}(x) \! :
\nonumber \\
&& \phantom{XXXXXXX}
- : \! \psi^{\dagger}_{R,\sigma}(x) i \frac{d}{dx}
\psi^{\mathop{\phantom{\dagger}}}_{R,\sigma}(x) \! : \biggr]
 \nonumber \\
 & + &  \frac{g}{2}
  : \! \psi^{\dagger}_{k,\sigma}(x)
  \psi^{\mathop{\phantom{\dagger}}}_{k,\sigma}(x) \! :
  : \! \psi^{\dagger}_{l, -\sigma}(x)
  \psi^{\mathop{\phantom{\dagger}}}_{l, -\sigma}(x) \! : \nonumber \\
 & + &  \, g \, : \! \psi^{\dagger}_{R,\sigma}(x)
 \psi^{\mathop{\phantom{\dagger}}}_{L,\sigma}(x) \psi^{\dagger}_{L,-\sigma}(x)
 \psi^{\mathop{\phantom{\dagger}}}_{R,-\sigma}(x) \! : \biggr\} , \label{TL}
\end{eqnarray}
%%%%%%%%%%%%%%%%%%%%%%%%%%%%%%%%%%%%%%%%
and coupled to a spin-$\frac{1}{2}$ impurity by
%%%%%%%%%%%%%%%%%%%%%%%%%%%%%%%%%%%%%%%%
\begin{equation}
{\cal H}_{el-imp} = \lambda_{kl} : \! \psi^{\dagger}_{k,\sigma}(0)
\mbox{$\frac{1}{2}$}
\mbox{\boldmath $\sigma$}^{\mathop{\phantom{\dagger}}}_{\sigma \mu}
\psi^{\mathop{\phantom{\dagger}}}_{l,\mu}(0) \! : \cdot \mbox{\boldmath $S$}.
\label{el-imp}
\end{equation}
%%%%%%%%%%%%%%%%%%%%%%%%%%%%%%%%%%%%%%%
Here $\psi_{L/R,\sigma}(x)$ are the left/right moving components
of the electron field $\Psi_{\sigma}(x)$, expanded about the
Fermi points $\pm k_F$, and we implicitly sum over repeated indices for
spin $\sigma, \mu = \uparrow, \downarrow$ and handedness $k, l = L, R$.
Normal ordering :: is carried out w.r.t. the filled Dirac sea.
The couplings $\lambda_F \equiv \lambda_{LL} \equiv \lambda_{RR}$ and
$\lambda_B \equiv \lambda_{LR} \equiv \lambda_{RL}$ are the amplitudes for
forward and backward electron scattering off the impurity
$\mbox{\boldmath $S$}$, respectively. For the physically relevant case
$\lambda_F = \lambda_B$ {\em (Kondo interaction)},
${\cal H}_{TL} + {\cal H}_{el-imp}$ contains the long-wavelength
physics of a small-$U$ Hubbard chain off half-filling, and coupled
to a single spin-$\frac{1}{2}$ impurity.
Then $g = Ua/2\pi$ and $v_F = 2at \sin a k_F$, with $U$ and $t$ the
usual Hubbard parameters and $a$ the lattice spacing.

The bulk Hamiltonian ${\cal H}_{TL}$ can be written on diagonal Sugawara
form \cite{Goddard}, using the charge and spin currents
%%%%%%%%%%%%%%%%%%%%%%%%%%%%%%%%%%
\begin{mathletters}
\begin{eqnarray}
j_{L/R}(x) & = & \cosh\theta : \! \psi^{\dagger}_{L/R,\sigma}(x)
\psi^{\mathop{\phantom{\dagger}}}_{L/R,\sigma}(x)
\! : \nonumber \\
& + & \sinh\theta : \! \psi^{\dagger}_{R/L,\sigma}(x)
\psi^{\mathop{\phantom{\dagger}}}_{R/L,\sigma}(x) \! : , \\
\mbox{\boldmath $J$}_{L/R}(x) & = & : \! \psi^{\dagger}_{L/R,\sigma}(x)
\mbox{$\frac{1}{2}$}
\mbox{\boldmath $\sigma$}^{\mathop{\phantom{\dagger}}}_{\sigma \mu}
\psi^{\mathop{\phantom{\dagger}}}_{L/R,\mu}(x) \! : ,
\end{eqnarray}
\end{mathletters}
%%%%%%%%%%%%%%%%%%%%%%%%%%%%%%%%%%%
with $\tanh2\theta = g/(v_F+g)$. Dropping a marginally
irrelevant term $-(g/\pi) \mbox{\boldmath $J$}_L \cdot
\mbox{\boldmath $J$}_R$, one obtains the critical bulk
Hamiltonian
%%%%%%%%%%%%%%%%%%%%%%%%%%%%%%%%%%%
\begin{equation}
{\cal H}^{*}_{TL} = \int \! dx
\biggl\{ \frac{v_c}{8\pi} \! : \! j_l(x)j_l(x) \! :
 + \, \frac{v_s}{6\pi} \! : \! \mbox{\boldmath $J$}_l(x) \cdot
 \mbox{\boldmath $J$}_l(x) \! : \biggr\} .
\label{Sugawara}
\end{equation}
%%%%%%%%%%%%%%%%%%%%%%%%%%%%%%%%%%%
The spin and charge separation in (\ref{Sugawara}) yields two
dynamically independent theories, each Lorentz invariant with a
characteristic velocity, $v_c = v_F (1 + 2g/v_F)^{\frac{1}{2}}$ and
$v_s = v_F - g$. The currents $j_l(x)$ and {\boldmath $J$}$_l(x)$ satisfy
the (level-2) $U(1)$ and (level-1) $SU(2)_1$ Kac-Moody algebras, respectively,
i.e. ${\cal H}^{*}_{TL}$ is invariant under the chiral symmetry
$U(1)_L \times U(1)_R \times SU(2)_{1,L} \times SU(2)_{1,R}$.

To cast the problem on a form where $BCFT$ applies, we use a
representation where the impurity location $x=0$ defines a {\em
boundary}. For this purpose we confine the system to the finite interval
$x \in [-\ell,\ell]$, fold it in half to $[0,\ell]$, identify the
two points $x=\pm \ell$, and introduce new currents for $x \ge 0$:
$j_{L/R}^1(x) \equiv j_{L/R}(x)$, $j_{L/R}^2(x) \equiv j_{R/L}(-x)$,
and analogously for {\boldmath $J$}$_l(x)$. We thus arrive at a
representation with doubled degrees of freedom on half the interval.
In 2D Euclidean space-time $\{z=v\tau + ix\}$, with $v=v_c$ ($v_s$)
for charge (spin), we interpret the time axis as a boundary where
%%%%%%%%%%%%%%%%%%%%%%%%%%%%%%
\begin{equation}
j_{L/R}^1(\tau,0) = j_{R/L}^2(\tau,0), \ \ \
\mbox{\boldmath $J$}_{L/R}^1(\tau,0) = \mbox{\boldmath $J$}_{R/L}^2(\tau,0).
\label{bc}
\end{equation}
%%%%%%%%%%%%%%%%%%%%%%%%%%%%%
By analytic continuation, this is equivalent to a chiral (left-handed)
theory on $[-\ell,\ell]$. The Hamiltonian then takes the form (\ref{Sugawara}),
but with the sum over handedness replaced by a sum over channels $1$ and $2$ of
left-handed currents only.

It is instructive to first study the case of only
{\em forward scattering} off the impurity, i.e.
$\lambda_{LR}=\lambda_{RL}$ = 0 in (\ref{el-imp}):
%%%%%%%%%%%%%%%%%%%%%%%%%%%%%%%
\begin{equation}
{\cal H}_{F} = \lambda_F [ \mbox{\boldmath $J$}_L^1(0) +
\mbox{\boldmath $J$}_L^2(0) ] \cdot \mbox{\boldmath $S$}.
\label{newFORWARD}
\end{equation}
%%%%%%%%%%%%%%%%%%%%%%%%%%%%%%%
As the two currents are coupled via $\mbox{\boldmath $S$}$, ${\cal H}_{F}$
breaks the $SU(2)_1 \times SU(2)_1$ symmetry of ${\cal H}^*_{TL}$ down to
the diagonal level-2 subalgebra $SU(2)_2$ spanned by
$\mbox{\boldmath $J$}(x) = \mbox{\boldmath $J$}_L^1(x)
+ \mbox{\boldmath $J$}_L^2(x)$. To adopt to this fact we use the
Goddard-Kent-Olive construction \cite{GKO} to write the spin part
of the Hamiltonian as a sum of an $SU(2)_2$ Sugawara Hamiltonian
and an Ising model. We can then absorb ${\cal H}_F$ into ${\cal H}^{*}_{TL}$
by the canonical transformation
$\mbox{\boldmath $J$}(x) \rightarrow \mbox{\boldmath $J$}'(x)
\equiv \mbox{\boldmath $J$}(x) + \mbox{\boldmath $S$} \delta (x) $,
$\mbox{\boldmath $J$}'(x)$ being the spin current
of electrons {\em and} impurity. The impurity thus disappears from the
Hamiltonian, and as a consequence (\ref{bc}) gets ``renormalized.''
This new, renormalized boundary condition is most easily defined by the
{\em selection rule} that prescribes how the $U(1) \times U(1)$, $SU(2)_2$,
and Ising degrees of freedom recombine at the boundary {\em after} the shift
$\mbox{\boldmath $J$}(x) \rightarrow \mbox{\boldmath $J$}'(x)$.

Consider first the unperturbed problem with $\lambda_F=0$. The charge
eigenstates organize into a product of two $U(1)$ conformal towers, one
for each channel, and labeled by two integer quantum numbers $(Q,
\Delta Q)$, the sum and difference of net charge in the two channels
(w.r.t. the groundstate).
These eigenstates are in 1-1 correspondence to the scaling operators
in the charge sector, of dimensions
\begin{equation}
\label{c-dim}
 \Delta_c = \frac{1}{4} \left( q_1^2 + q_2^2 \right) + N_c ,
\end{equation}
with
\begin{equation}
\label{q1q2}
q_i = Q \frac{e^{\theta}}{2} - (-1)^{i} \Delta Q \frac{e^{-\theta}}{2}
\end{equation}
and $N_c \in {\Bbb N}$.
Similarly, the eigenstates in the $SU(2)_2$ and $Ising$
sectors appear in conformal towers labeled by the spin quantum numbers
$j=0,\frac{1}{2},1$, and the Ising primary fields
$\phi=\openone,\sigma,\epsilon$, respectively:
\begin{eqnarray}
\Delta_{S} & = & \frac{1}{4}j(j+1) + N_{S}, \label{S-dim} \\
\Delta_{Ising} & = & 0 (\openone), \frac{1}{16} (\sigma), \label{Ising-dim}
\frac{1}{2} (\epsilon) + N_{Ising},
\end{eqnarray}
where $N_S, N_{Ising} \in {\Bbb N}$.
The complete set of conformal towers is accordingly labeled by $(Q,
\Delta Q, j, \phi)$ and the spectrum of scaling dimensions is $\Delta =
\Delta_c + \Delta_S + \Delta_{Ising}$. The selection rule for
combining quantum numbers can be extracted from comparison with
{\em Bethe Ansatz} results for the Hubbard model \cite{Woynarovich}
(of which ${\cal H}_{TL}$ is the long-wavelength effective theory),
and one finds: $(j,\phi) =
(0,\openone)$ or $(1,\epsilon)$ for $Q, \frac{1}{2}(Q+\Delta Q)$ even;
$(0,\epsilon)$ or $(1,\openone)$ for $Q$ even and $\frac{1}{2}(Q+\Delta Q)$
odd; $(\frac{1}{2}, \sigma)$ for $Q$ odd.

When $\lambda_F \neq 0$ we absorb ${\cal H}_F$ into ${\cal
H}^{*}_{TL}$ by redefining the spin current as that of
electrons {\em and} impurity. Effectively, this adds an extra
spin-$\frac{1}{2}$ degree of freedom to the $SU(2)_2$ towers, which,
as a result, get shifted according to the conformal
field theory {\em fusion rules}: $j=0
\rightarrow \frac{1}{2}$, $\frac{1}{2} \rightarrow 0$ or $1$,
$1 \rightarrow \frac{1}{2}$.
The selection rule describing the new content of possible boundary
scaling operators is obtained by applying fusion {\em twice} to the
previous selection rule \cite{AL}. This gives for forward scattering:
$(j,\phi) = (0$ or $1$, $\openone$ or $\epsilon)$ for $Q$ even;
$(\frac{1}{2},\sigma)$ for $Q$ odd.

The low-temperature thermodynamics is now governed by the
{\em leading correction-to-scaling boundary operator (LCBO)}.
As this must preserve all symmetries of ${\cal H}^*_{TL} + {\cal H}_{F}$,
the forward scattering selection rule together with invariance under
chiral $U(1)$, $SU(2)_2$, and channel exchange $(1 \leftrightarrow 2)$,
imply a unique $LCBO$ given by the first descendant in the $j=1$
tower: $\mbox{\boldmath $J$}_{-1} \cdot \bbox{\phi}$.
This is the same {\em LCBO} that drives
critical scaling in the two-channel Kondo effect for noninteracting
electrons \cite{AL}. Specifically, the impurity contributions to
the specific heat $\delta C $ and spin susceptibility $\delta \chi $
are given to leading order by
%%%%%%%%%%%%%%%%%%%%%%%%%%%%%%%%%%%%%%%%%%
\begin{mathletters}
\begin{eqnarray}
\delta C & = & \frac{\mu_F^2 9 \pi^2}{v_s^3}
 T \ln (\frac{1}{\tau_0T}), \label{Cimp} \\
\delta \chi & = & \frac{\mu_F^2 18}{v_s^3} \ln (\frac{1}{\tau_0T}),
\label{CHIimp}
\end{eqnarray}
\end{mathletters}
%%%%%%%%%%%%%%%%%%%%%%%%%%%%%%%%%%%%%%%%%%
as $T \rightarrow 0$. Here $\mu_F$ is the scaling field conjugate
to $\mbox{\boldmath $J$}_{-1} \cdot \bbox{\phi}$ and $\tau_0$ a short-time
cutoff. With the known bulk response for the Tomonaga-Luttinger model,
$C = \pi (v^{-1}_c + v^{-1}_s)T/3$ and $\chi = 1/2\pi v_s$ \cite{KY},
we predict a Wilson ratio
%%%%%%%%%%%%%%%%%%%%%%%%%%%%%%%%%%%%%%%%%%%%
\begin{equation}
%R_W = (\frac{\delta \chi}{\chi}) / (\frac{\delta C}{C}) =
R_W = \frac{\delta \chi / \chi}{\delta C / C} =
\frac{4}{3}(1+\frac{v_s}{v_c}).
\label{Wilson}
\end{equation}
%%%%%%%%%%%%%%%%%%%%%%%%%%%%%%%%%%%%%%%%%%%%
For $g \rightarrow 0$ \ ($v_c$, $v_s \rightarrow v_F$), this reduces to
the universal number 8/3 characterizing the usual two-channel
Kondo effect \cite{AL}.

Let us now include backward scattering off the impurity,
$\lambda_B \equiv \lambda_{LR} \equiv \lambda_{RL} \neq 0$. The corresponding
terms in (\ref{el-imp}) break the chiral $SU(2)$ {\em and} chiral $U(1)$
invariance of ${\cal H}^*_{TL}$. As a consequence $\Delta Q$ is no longer
restricted to zero, and the charge sector makes non-trivial contributions
to the content of scaling operators. The lowest dimension operator with
$\Delta Q \neq 0$ allowed by the forward scattering selection rule is
obtained from $(Q, \Delta Q, j, \phi) = (0, \pm 2, 0 , \openone)$, and has
dimension $\Delta = \frac{1}{2} e^{-2\theta} \le \frac{1}{2}$. Back scattering
is thus a relevant perturbation and drives the system to a new fixed point.
When the flows of $\lambda_F$ and $\lambda_B$ converge, this is the fixed point
for {\em Kondo scattering} in a Luttinger liquid.

To study this case we consider the bare Kondo interaction
%%%%%%%%%%%%%%%%%%%%%%%%%%%%%%%%%%%%%%%%%%%%%
\begin{equation}
{\cal H}_K = \lambda \sum_{k,l = L, R} : \! \psi^{\dagger}_{k,\sigma}(0)
\mbox{$\frac{1}{2}$}
\mbox{\boldmath $\sigma$}^{\mathop{\phantom{\dagger}}}_{\sigma \mu}
\psi^{\mathop{\phantom{\dagger}}}_{l,\mu}(0) \! : \cdot \mbox{\boldmath $S$},
\end{equation}
%%%%%%%%%%%%%%%%%%%%%%%%%%%%%%%%%%%%%%%%%%%
obtained from (\ref{el-imp}) by choosing $\lambda_{kl} = \lambda$, i.e.
$\lambda_F = \lambda_B = \lambda$.
With no e-e interaction ($g=0$ in (\ref{TL})) we have a free bulk
Hamiltonian ${\cal H}_0$ together with  ${\cal H}_{K}$.
Passing to a basis spanned by definite-parity fields
$\psi_{\pm,\sigma}(x) = [ \psi_{L,\sigma}(x) \pm \psi_{R,\sigma}(-x)]
/\sqrt{2}$, ${\cal H}_0 + {\cal H}_{K}$ transforms into a
two-channel theory, but with the impurity coupled to the electrons
in only one of the channels. This renormalizes to a local Fermi liquid
(like the ordinary 3D Kondo problem), with response
functions scaling analytically with temperature \cite{NozieresBlandin}.
However, a different approach must be used for the interacting problem since
${\cal H}^*_{TL}$ is non-local in this basis. Here we exploit the
expectation that {\em any} local impurity interaction, including the Kondo
interaction ${\cal H}_K $, can be substituted by a renormalized boundary
condition on the critical bulk theory \cite{AffleckTaniguchi}.
The equivalent selection rule defines
a fixed point, and by demanding that any associated $LCBO$ must
respect the symmetries of the problem {\em and} correctly reproduce the
non-interacting limit as $g \rightarrow 0$, the possible critical theories
can be deduced. (Note that a selection rule here defines a {\em boundary}
fixed point, and is valid for all values of the marginal bulk
coupling $g$. Hence, given a selection rule,
Fermi liquid scaling must emerge in the limit $g \rightarrow 0$.)

To have a generally applicable formalism we introduce a notation that
does not make an implicit relation between the two diagonalized charge
towers (as $Q$ and $\Delta Q$ do), and denote a combination of
conformal towers by $(C_1, D_1; \ C_2, D_2 ; \ j; \ \phi)$.
Hence $(C_i, D_i)$ replace $Q$ and $\Delta Q$, such that the scaling
dimensions in the charge sector are now given by (\ref{c-dim}), with
\begin{equation}
q_i = C_i \frac{e^{\theta}}{2} - (-1)^{i} D_i
\frac{e^{-\theta}}{2}
\end{equation}
replacing (\ref{q1q2}).
The corresponding states are seen to be
global $U(1)$ invariant if $q \equiv q_1+q_2=0$, and chiral $U(1)$ invariant if
$q= \Delta q \equiv q_1-q_2 =0$. This is consistent with our
previous notion of global and chiral $U(1)$ invariance in terms of $Q$
and $\Delta Q$, as the former selection rules implied the relation
$C_1=C_2$ and $D_1=D_2$.
The crucial point to realize is that $Q$ and $\Delta Q$ are not
sufficient to label {\em all} combinations of $U(1)$ conformal
towers, whereas $q$ and $\Delta q$ are well-defined for {\em any}
selection rule.
Hence, at the new fixed point, the signature of breaking chiral
$U(1)$ invariance is to allow operators with $\Delta q \neq 0$.
Global $U(1)$ invariance, on the other hand,
respected by ${\cal H}_K$, requires $q=0$. Together with invariance
under channel exchange $(1 \leftrightarrow 2)$, this leaves
only two possibilities for the charge part of the {\em LCBO}
\cite{FJlong}: (i) $C_1 = C_2 = 0$, $D_1 = D_2 =$ even
integer $\Rightarrow \Delta_c = \frac{1}{2} p^2  e^{-2\theta} + N_c$,
and (ii) $C_1 = - C_2 =$ even
integer, $D_1  = D_2 = 0$ $\Rightarrow \Delta_c = \frac{1}{2} p^2
e^{2\theta} + N_c$, with $p, N_c \in {\Bbb N}.$

The complete scaling dimensions are obtained by coupling the $SU(2)_2$
and Ising conformal towers to the pairs of $U(1)$ towers in $(i)$
and $(ii)$. Starting with the $SU(2)_2$ sector, the
$j=\frac{1}{2}$ tower is expelled by global SU(2) invariance. Turning to
the $j=1$ tower, the primary operator
$\bbox{\phi}$ is excluded by the same reason. The lowest-dimension
$SU(2)_2$ singlet operator from this tower
is $\mbox{\boldmath $J$}_{-1} \cdot \bbox{\phi}$. However, this is the same
operator that drives critical scaling in the forward scattering problem.
It produces a diverging impurity susceptibility as $T
\rightarrow 0$, in conflict with the known Fermi
liquid scaling in the $g \rightarrow 0$ limit. The $j=1$ tower is therefore
expelled and the only contribution from the $SU(2)_2$ sector is
the identity and its descendants. Next we note that
no relevant scaling operators are allowed, since at $g=0$ the fixed point is
known to be stable, being that of the ordinary Kondo problem. As $g$ is
the {\em only} tunable parameter in ${\cal H}_{TL}^*$
(with a renormalized boundary condition replacing
$H_K$), this is true also for $g \neq 0$ since otherwise the theory would
become noncritical. Hence, starting with $(i)$ and $p=0$, only $\openone$
from the Ising sector is permissible, as any other choice would produce a
relevant operator. For $p=1$, all choices lead to relevant operators,
whereas for $p \geq 2$ the converse is true.
Summarizing, the possible couplings of $SU(2)_2$ and Ising towers to the
$U(1)$ towers selected by $(i)$ yield the following candidate {\em LCBO}
dimensions:
%%%%%%%%%%%%%%%%%%%%%%%%%%%%%%%%%%%
\begin{equation}
\Delta_{LCBO} = 1, \ \frac{1}{2} p^2 e^{-2\theta} +
\{0, \frac{1}{16}, \frac{1}{2} \},
\label{candI}
\end{equation}
%%%%%%%%%%%%%%%%%%%%%%%%%%%%%%%%%%%%
with $p \in {\Bbb N} +2$. Here $\Delta_{LCBO} = 1$ is the dimension of the
first $U(1)$ Kac-Moody descendants $j^{1,2}_L$, allowed by the
broken particle-hole symmetry of the underlying lattice model.
Turning to $(ii)$, and employing the same reasoning as above, one finds
a second class of possible  {\em LCBO} dimensions:
%%%%%%%%%%%%%%%%%%%%%%%%%%%%%%%%%%%%
\begin{equation}
\Delta_{LCBO} = 1, \ \frac{1}{2} e^{2\theta} + \frac{1}{2},
\ \frac{1}{2} p^2 e^{2\theta} + \{0, \frac{1}{16}, \frac{1}{2} \},
\label{candII}
\end{equation}
%%%%%%%%%%%%%%%%%%%%%%%%%%%%%%%%%%%
and with $p$ as above.

Each entry in (\ref{candI}) and (\ref{candII}) defines an effective
scaling Hamiltonian ${\cal H}_{scaling} = {\cal H}^*_{TL} + \mu {\cal
O}(0)$, with ${\cal O}(0)$ the corresponding $LCBO$ conjugate to the
scaling field $\mu$. Using ${\cal H}_{scaling}$, the finite-size corrections
at the fixed point can be calculated perturbatively in $\mu$,
and by treating temperature as an inverse length, the
corrections to the bulk thermodynamics due to the impurity are accessible
via finite-size scaling. Given (\ref{candI}) and
(\ref{candII}), and requiring Fermi-liquid scaling for the impurity
specific heat $\delta C$ and susceptibility $\delta \chi $ as $g
\rightarrow 0$, we find that there are only {\em two} possible types of
critical behavior. When $\Delta_{LCBO} = 1$ or $\Delta_{LCBO} > \frac{3}{2}$
Fermi-liquid scaling persists for $g \neq 0$, whereas a
non-Fermi liquid behavior emerges when
$\Delta_{LCBO} = \frac{1}{2} (e^{2\theta} + 1)$:
%%%%%%%%%%%%%%%%%%%%%%%%%%%%%%%%%%%%
\begin{mathletters}
\begin{eqnarray}
\delta C & = & c_1((1/K_{\rho})-1)^2T^{(1/K_{\rho})-1} + c_2 \, T, \\
\delta \chi & = & c_3 \, T^0,
\end{eqnarray}
\label{anomalous}
\end{mathletters}
%%%%%%%%%%%%%%%%%%%%%%%%%%%%%%%%%%%%%
as $T \rightarrow 0$. Here $K_{\rho} = (1+2g/v_F)^{-1/2}$ and $c_{1,2,3}$
are amplitudes depending on the scaling fields and velocities.
The $LCBO$ driving the anomalous scaling in (\ref{anomalous}) is
given by the composite operator
${\cal O}_{LCBO} = [V^1_{2,0} \times V^2_{-2,0} +
V^1_{-2,0} \times V^2_{2,0}] \times \epsilon$ where
$V^i_{C,D}$ is a $U(1)$ primary (vertex) operator in channel $i$,
and $\epsilon$ the Ising energy density.
This scaling (\ref{anomalous}) agrees exactly with that proposed by
Furusaki and Nagaosa \cite{FN}, in support of a non-Fermi liquid scenario.
However, a simplified model (neglecting backward spin diagonal and forward
spin off-diagonal Kondo scattering) suggests that in fact the other scenario
(Fermi liquid) may be realized \cite{SI}. Note that in none of the two cases
does the e-e interaction influence $\delta \chi$: the impurity remains
completely screened for $g\neq 0$.

In summary, we have shown that the symmetries of the problem
restrict the possible critical theories to {\em either} a
local Fermi liquid (as for free 3D electrons) {\em or}
a non-Fermi liquid with thermodynamic response as in (\ref{anomalous}).
The $BCFT$ approach as presented here is quite general and can be used to
derive the finite-size energy spectrum at the non-Fermi liquid fixed
point, as well as transport properties.
Details will be published elsewhere \cite{FJlong}.

We thank  A. A. Nersesyan and E. Wong for many illuminating discussions
on this topic. We are also indebted to I. Affleck, M. P. M. den Nijs,
D. Kim, and A. W. W. Ludwig for useful comments and suggestions.
The authors acknowledge support from NSF grants DMR-9205125 and
DMR-91-120282 (P. F.), and a grant from the Swedish Natural Science
Research Council (H. J.).

%%%%%%%%%%%%%%%%%%%%%%%%%%%%%%%%%%%%%%%%%%%%%%%%%%%%%%%%%%%%

%%%%%%%%%%%%%%%%%%%%%%%%%%%%%%%%%%%%%%%%%%%%%%%%%%%%%%%%%%%%


\begin{references}
\bibitem{Chiral} The important exception is a chiral 1D system,
such as edge excitations in a quantum Hall liquid. See e.g. \cite{Timp}.
\end{references}
\end{document}